\documentclass[reprint,superscriptaddress,amsmath,amssymb,aps,prb,floatfix]{revtex4-2}
\usepackage[english]{babel}
\usepackage{graphicx}
\usepackage{xcolor}
\usepackage{latexsym}
\usepackage{textcomp}
\usepackage{soul}
\usepackage{siunitx}
\usepackage{upgreek}
\usepackage[version=4]{mhchem}
\usepackage[hidelinks]{hyperref}
\newcommand{\equalcontribution}{These authors contributed equally to this work.}

\begin{document}

\title{Effect of pulse duration on current-induced selective oxygen migration in high-\texorpdfstring{T\textsubscript{c}}{Tc} superconductors}  

\author{Fridrich Egyenes}
\thanks{\equalcontribution}
\affiliation{Institute of Electrical Engineering, Slovak Academy of Sciences, Bratislava 841 04, Slovakia}
\affiliation{Faculty of Physics, University of Vienna, 1090 Vienna, Austria}
\author{Daniel Stoffels}
\thanks{\equalcontribution}
\affiliation{Experimental Physics of Nanostructured Materials, Q-MAT, Department of Physics, Universit\'{e} de Li\`{e}ge, B-4000 Sart Tilman, Belgium}
\author{Stefan Marinkovi\'{c}} 
\affiliation{Experimental Physics of Nanostructured Materials, Q-MAT, Department of Physics, Universit\'{e} de Li\`{e}ge, B-4000 Sart Tilman, Belgium}
\affiliation{National High Magnetic Field Laboratory, Los Alamos National Laboratory, 87545 Los Alamos, NM, USA.}
\author{Bernd Aichner}
\affiliation{Faculty of Physics, University of Vienna, 1090 Vienna, Austria}
\author{Huidong Li}
\affiliation{Institut de Ciència de Materials de Barcelona, ICMAB-CSIC, Campus UAB, 08193 Bellaterra, Spain}
\author{Anna Palau}
\affiliation{Institut de Ciència de Materials de Barcelona, ICMAB-CSIC, Campus UAB, 08193 Bellaterra, Spain}
\author{Milan \v{T}apajna} 
\affiliation{Institute of Electrical Engineering, Slovak Academy of Sciences, Bratislava 841 04, Slovakia}
\author{Wolfgang Lang}
\email{wolfgang.lang@univie.ac.at}
\affiliation{Faculty of Physics, University of Vienna, 1090 Vienna, Austria}
\author{Alejandro V. Silhanek}
\affiliation{Experimental Physics of Nanostructured Materials, Q-MAT, Department of Physics, Universit\'{e} de Li\`{e}ge, B-4000 Sart Tilman, Belgium}

\date{\today}

\begin{abstract}
High current densities can induce the directional diffusion of atoms in metallic films. In YBa$_2$Cu$_3$O$_{7-\delta}$ (YBCO), this electromigration process selectively acts on oxygen atoms lying in the Cu-O chains, permitting to vary the oxygen concentration in a targeted spot of high current density. This approach has proven successful in mapping the phase diagram of the material as a function of carrier concentration or as a way to manufacture memristive devices owing to its reversibility under small bipolar excitations. Thus far, most of the investigations have been limited to pulsed excitation with current/voltage pulses on a millisecond or longer scale, for which thermal effects undeniably influence the process. In the present work, we explore the impact of pulse length $\delta t$ on the onset current of electromigration, $I_{\text{EM}}$, of YBCO bridges, covering the range from 200 ns to 1 ms. As $\delta t$ decreases below $\sim 10$ µs, $I_{\text{EM}}$ exhibits a rapid increase. Analytical and numerical estimates of the local temperature show that as pulses shorten, the temperature decreases, making the electromigration process more athermal. These findings are relevant for the operation of memristors and should be taken into account when describing the effects of thermomagnetic instabilities in thin films.
\end{abstract}

\maketitle

\section{Introduction}

Electromigration is a process by which charge carriers transfer their linear momenta to atoms, resulting in a slow and directional atomic diffusion process along the current flow. The atomic drift can be downstream or upstream, depending on the sign of what is known as the effective valence $z^*$ of the atomic species in question \cite{Lloyd1997}. Particularly favourable conditions for triggering electromigration arise in environments characterised by static disorder (defects, interfaces, grain boundaries, etc.) or dynamic disorder (phonons) due to elevated temperatures. Electromigration rarely occurs in isolation; it coexists with a variety of sometimes competing phenomena stemming from temperature gradients (thermomigration), mechanical stress gradients (stress migration), or vacancy concentration gradients (diffusion) \cite{hoffmann_vogel_electromigration_2017}. The complex interplay of these mechanisms results in counterintuitive behaviours such as atom displacement in regions of low current density \cite{Lombardo2019a}, relaxation after driving forces are removed \cite{Lloyd, Baumans2017, Collienne2022}, or even partial healing if the driving force reverses polarity \cite{Kozlova2013}. 

Both pulsed and bipolar electrical signals have been recognised as highly promising approaches to prolong the lifespan of integrated circuits susceptible to failure due to electromigration \cite{PIERCE1997}. Initially, it was assumed that electromigration damage under pulsed conditions would depend solely on the pulse duration $\delta t$. However, it was later demonstrated that the repetition rate $\tau$ plays a significant role, with different regimes identified as a function of the duty cycle $\delta t/\tau$ or frequency $1/\tau$ \cite{English-72,Schoen,PIERCE1997}. Pulsed electromigration has been shown to be an effective strategy for controlling this inherently stochastic process in Au nanowires under 10 µs voltage pulses \cite{Hayashi_2008} or much longer pulses \cite{Wu}, as well as in Al \cite{Lombardo2019a}, and even in refractory materials such as Nb \cite{Stefan-2023}. It is important to note that in all instances, the pulse-on-time exceeded the thermal relaxation time of the investigated nanowire (typically on the sub-µs scale), and thus the electromigration process was mostly thermally assisted.

Recently, the interest of the scientific community has expanded from metallic circuits relevant to information technology to investigating the electromigration of high critical-temperature superconductors (HTSC), particularly copper oxides. The motivation behind this fact lies in the high mobility of oxygen atoms and the inherent disorder of oxygen distribution in these materials, along with the rich diversity of electronic phases that emerge with varying oxygen content \cite{Roadmap}. Targeted current-induced oxygen migration has enabled an all-electrical tuning of the material's properties \cite{Marinkovic2020}, mapping the phase boundary of YBa$_2$Cu$_3$O$_{7-\delta}$ (YBCO) films \cite{Trabaldo2022}, adjusting the response of superconducting weak links \cite{Trabaldo2021}, stimulating oxygen ordering \cite{Marinkovic2023} and replenishing oxygen from where it was removed \cite{Marinkovic-2024}. In addition, there is a genuine interest in understanding the response of HTSC superconducting devices (cables, single-photon detectors, fault current limiters, etc.) to pulsed high-current excitations. Although the superconducting state remains immune to the electromigration problem, high-current excitation able to reach the dissipative state can lead to migration of other atomic species causing irreversible damage \cite{Baumans2019}. A possible way to ensure more resilient devices is to shorten the on-time of the pulse, thus reducing the delivered energy and, consequently, the local temperature rise. Although pulsed electrical measurements in HTSC have been amply investigated in the vicinity of the superconducting state \cite{GUPTA-1993,Puica_2004,Puica-2019}, little is known about the threshold current $I_{\text{EM}}$ beyond which atomic diffusion is triggered in the normal state, or in ambient conditions, nor its dependence on pulse length.

The present work investigates the upper limiting current $I_{\text{EM}}$ that a YBCO microbridge can stand, in the normal state, without undergoing major structural modifications. The focus is on single isolated pulses with nearly zero duty cycle at room temperature and atmospheric pressure. We show that $I_{\text{EM}}$ sharply increases as the pulse length decreases below 10 µs. This behavior is qualitatively explained through analytical and numerical models, showing that the electromigration process becomes progressively more athermal as the pulse length is reduced. These findings may be relevant to the investigation of other oxide films, such as manganite, subjected to extreme current densities required for magnetic domain wall displacement induced by the spin-torque effect, and they complement recently reported voltage-pulse experiments in YBCO memristive devices \cite{Rouco}.  

\section{Experimental}

\begin{figure}[!ht]
    \centering
    \includegraphics[width=\linewidth]{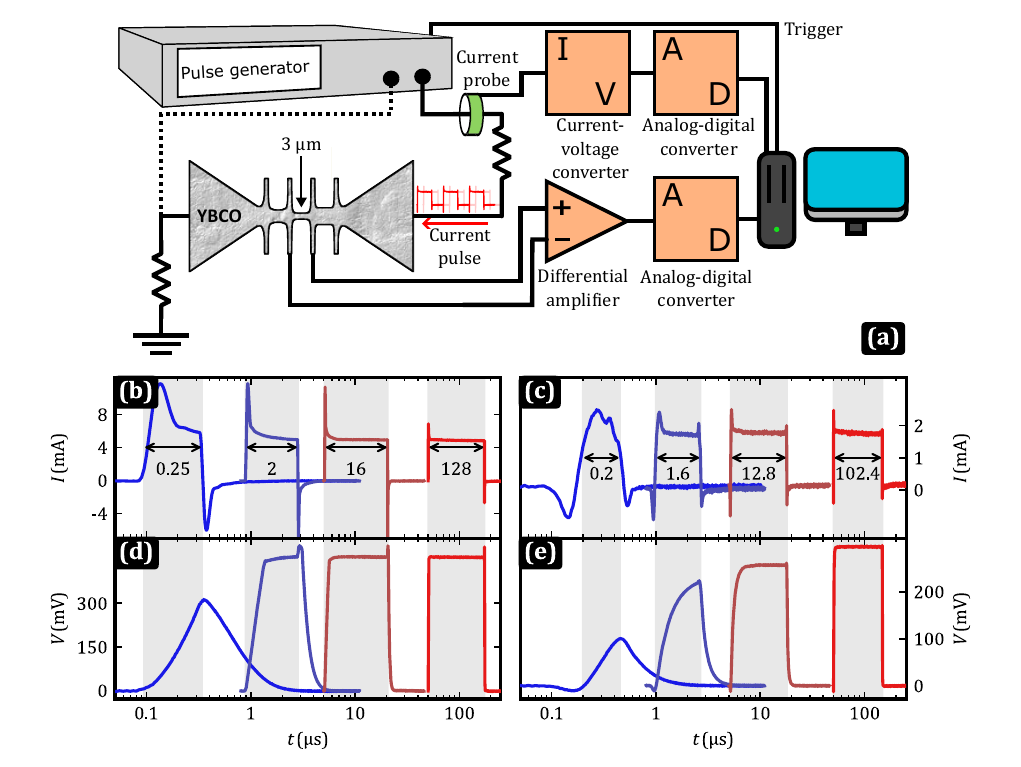}
    \caption{
    \label{fig:setup}
    (a) Schematics of the measurement setup. Two different pulse generators are used. One contact of the sample is always connected to the pulse generator's output, whereas the other contact is either connected to ground (with the unipolar pulse generator) or to the bipolar pulse generator, as pictured by the full or dotted lines on the left of the sample, respectively. The pulse generator applies a voltage along the samples and the current thus circulates horizontally. The vertical voltage leads allow for a four-point measurement of the constriction resistance.
    Panels (b)-(e) show the pulse shape evolution for the unipolar (b,d) and bipolar (c,e) pulse generator output current (b,c) and four-point voltage across the constriction (d,e) as a function of the pulse length. The distortion of the voltage pulse is apparent for the shorter pulses, whereas the current pulses remain less affected. The pulse length is indicated in panels (b) and (c) in µs.
    } 
\end{figure}

The investigated samples consist of 50 nm thin films of YBCO epitaxially deposited on \ce{LaAlO3} single crystals by pulsed laser deposition. The films are patterned by photolithography and ion-beam etching into a triple-constriction design, with the constrictions connected in series \cite{Marinkovic2020,Marinkovic-2024}. The outer constrictions have a width three times the width of the inner one, localizing the nucleation point of electromigration to the inner constriction by means of the current crowding effect. The layout of the sample is sketched in Fig.~\ref{fig:setup}(a).

All pulsed current-induced oxygen migration experiments are performed at room temperature. The samples are wirebound to a printed circuit board holder and mounted in a shielding metal box for the experiment. The investigation of electromigration induced by short current pulses was conducted using the electric circuit illustrated in Fig.~\ref{fig:setup}(a). Two types of voltage-pulse generators were used: (a) a programmable unipolar pulse generator (HP 8114 A 100V/2A) which can provide voltage pulses with amplitude from 2 to 100 V and pulse duration $\delta t$ between 10 ns and 10 ms in either positive or negative polarity; and (b) a programmable pulse generator with two symmetric-to-ground output voltages (Avtech AVR-7B-B-PN) supplied simultaneous constant-voltage pulses with amplitudes ranging from ±1 V to ±100 V and a maximum current of 2 A, and the adjustable pulse duration between 100 ns and 100 µs. To show the minor effect of the used setup, we ensured to use both setups under comparable conditions: the minimum current in the load was $\approx $ 0.5 mA, and the minimum used pulse duration was as short as 100 ns. Two identical low-inductance resistors were used to limit the current supplied to the sample and thus ensure nearly constant-current conditions. The current was monitored using a high-speed current clamp and a current-to-voltage converter (Tektronix P6302 and AM503). A direct comparison of both pulse-generation setups under matched experimental conditions is provided in Appendix~\ref{sec:app_setup} (Fig.~\ref{fig:app_setup}).

The requirement for a four-terminal electrical measurement using high-impedance voltage probes to eliminate contact effects makes the conventional asymmetric $50\, \Omega$ technique largely impractical. To address this, the two voltage contacts of the sample were connected to the differential input channels of a fast preamplifier (LeCroy DA1855A), which features a high common-mode rejection ratio (CMRR) and rise/fall times of 3.5 ns. Although the preamplifier provides a $\mathrm{CMRR}=10^5:1$ for signal frequencies up to 100\,kHz, its rejection ratio decreases $\sim 1/ f$ at higher frequencies. Consequently, using symmetric excitation with a virtual ground level between the voltage contacts significantly reduces common-mode voltage, enabling the detection of short voltage pulses. Finally, the signals were digitized with 16-bit resolution using fast analog-to-digital converters operating at $2 \times 10^8$ samples/s (GaGe Razor 1642). A custom Python program was used to control the measurement protocol.

The measurement procedure follows a systematic approach in which a high-voltage pulse is applied to the load, and the resulting circulating current, limited by the load resistance, is measured with a current clamp. During the high-current pulse $I_{\text{pulse}}$, the resistance of the sample ($R_{\text{pulse}}$) is measured. Subsequently, in order to assess stoichiometric changes in the sample, its resistance  ($R_{\text{min}}$) is measured during a longer, low-amplitude current pulse, thus unaffected by Joule heating. This last value is stored and compared with the pristine value $R_{0}$. Then, the amplitude of the applied voltage pulse is incrementally increased by 20 mV, and the process is repeated until $R_{\text{min}}$ increases by 3\,\% ($R_{\text{min}}=1.03R_{0}$), indicating the onset of oxygen migration. At this point, the corresponding current is recorded as $I_{\text{EM}}(\delta t)$. 

Figure~\ref{fig:setup}(b-e) shows representative examples of current and voltage pulses of different pulse lengths $\delta t$ obtained with the unipolar (b,d) and bipolar (c,e) pulse generators. Long pulses down to $\sim 2$ µs retain their square shape but get progressively deformed for shorter pulses, as a consequence of the relatively high impedance of the load (total 2-wire resistance during the measurements was $ 5-6\,\mathrm{k \Omega}\ $). Then, the pulse current amplitude and device resistance were determined as follows: (i) the current through the load and the voltage drop across the device were recorded simultaneously; (ii) the current amplitude ($I_{\text{pulse}}$) was obtained by integrating the measured pulse current signal and normalizing it to the applied pulse length; (iii) the voltage amplitude ($V_{\text{pulse}}$) was extracted analogously by integrating the measured voltage drop signal and averaging it to the same pulse duration length; (iv) the pulse resistance ($R_{\text{pulse}}$) was then calculated from $V_{\text{pulse}}$ and $I_{\text{pulse}}$. To evaluate the accuracy of this approach, we conducted a control experiment in which the YBCO sample was replaced by resistors of known value, keeping the overall circuit load similar. The extracted amplitudes were then analyzed against the known resistance. The maximum deviation from the known resistance is about 15 \% for the unipolar pulse generator, and $\sim$ 2 \% for the bipolar pulse generator.
In the case of $R_{\text{min}}$ evaluation, we took advantage of long pulses, where the ringing effects were minimal, allowing for clear plateaus to be observed over a significant range of the collected data.

\section{Numerical Model\label{sec:3}}
In order to assess oxygen redistribution in the devices during electromigration, we perform time-dependent finite-element simulations using the proprietary software suite COMSOL Multiphysics\textsuperscript{\textregistered} \cite{comsol}. Multiple physics are solved concurrently in a fully coupled way: electric currents (to obtain the current-density distribution and device resistance), heat transfer in solids (to account for Joule heating), and oxygen transport driven both by chemical-potential gradients and by electromigration forces \cite{hoffmann_vogel_electromigration_2017}.

The key constitutive relation for the material is the normal-state resistivity, which depends on both temperature and oxygen content,
\begin{equation}
    \rho(T,\delta) = \rho_{300\mathrm{K}}(\delta)\,\bigl[1 + \alpha_T (T - T_0)\bigr],
    \label{eqn:rho_law}
\end{equation}
where \(\rho_{300\mathrm{K}}(\delta)\) is the in-plane resistivity at \(T_0 = 300\,\mathrm{K}\), and its dependence on the local oxygen content (or, equivalently, on \(\delta\) in YBa\(_2\)Cu\(_3\)O\(_{7-\delta}\)) is taken from Ref.~\cite{Semba2001}.

The overall simulation strategy is largely inspired by the procedure developed in Ref.~\cite{Collienne2022}, but major adaptations are required to capture the short-pulsed nature of the present protocol. We therefore use a transient study with current pulses of varying duration \(\delta t\). For each pulse, the total simulated time is chosen as twice the pulse on-time to allow for oxygen redistribution while the system remains hot after the pulse. In addition, we include finite rise and fall times of \(10\%\) of the pulse duration to avoid numerical artefacts. 

The geometry is modeled as a 2D YBCO film representing the bridge, deposited on a 3D substrate. Two thermal boundary conditions are applied to the substrate: the back side is held at \(T_0 = 300\,\mathrm{K}\), while the top surface is subject to a convective heat-flux boundary condition (see annex \ref{sec:app_mesh}).

As a first calibration step, the pristine digital twin is adjusted to reproduce the experimentally measured cold resistance \(R_{\mathrm{0}}(T_0)\) at the base temperature \(T_0 = 300\,\mathrm{K}\). To this end, we apply a low readout current \(I_{\text{read}} = 100\,\mathrm{\upmu A}\) in the simulations---small enough to neglect self-heating---and tune the effective film thickness \(d\) until the simulated resistance \(R_{\text{sim}}(T_0)\) matches the experimental value. For a film of nominally 50 nm, for instance, we obtain $d=59$ nm. This calibrated model is then used as the starting point for the pulsed-current simulations, in which the spatio-temporal evolution of both temperature and oxygen concentration is tracked during and after each pulse.

A second calibration step is required for the thermal modeling, since many boundary conditions as well as material and interfacial properties are not precisely known and depend sensitively on the fabrication process. It has been shown that electromigration in these devices can induce long-range heating, which would, in principle, require simulating the entire assembly (chip, holder, wirebonds, etc.) in order to capture heat transfer accurately \cite{Marinkovic-2024}. Such a full-device model is prohibitively demanding in computational resources, and therefore, we restrict the geometry to a small substrate beneath the thin film and justify this reduction by a dedicated thermal calibration procedure. In the experiment, we have access to the device resistance during the long pulse, i.e. close to thermal equilibrium, and then tune the convective heat-transfer coefficient and the effective substrate thermal conductivity to match the simulated pulse resistance to the experimental value, thereby reproducing the overall Joule-induced temperature rise.

\section{Results and Discussion}

The present report investigates the dependence of the electromigration current on pulse length. Because the monopolar electromigration process induces cumulative, irreversible changes in the sample, it is ideal to conduct a single electromigration run per sample. This constraint requires a large number of {\it identical} samples; otherwise, comparisons of the results will be inaccurate due to inevitable variability in sample geometry and properties. To avoid this, we have chosen to perform the pulse-length analysis on the same device, limiting the electromigration process to small resistance changes (3\%). This approach ensures that the geometry and properties remain comparable to the initial structure.

\begin{figure}[!ht]
    \centering
    \includegraphics[width=\linewidth]{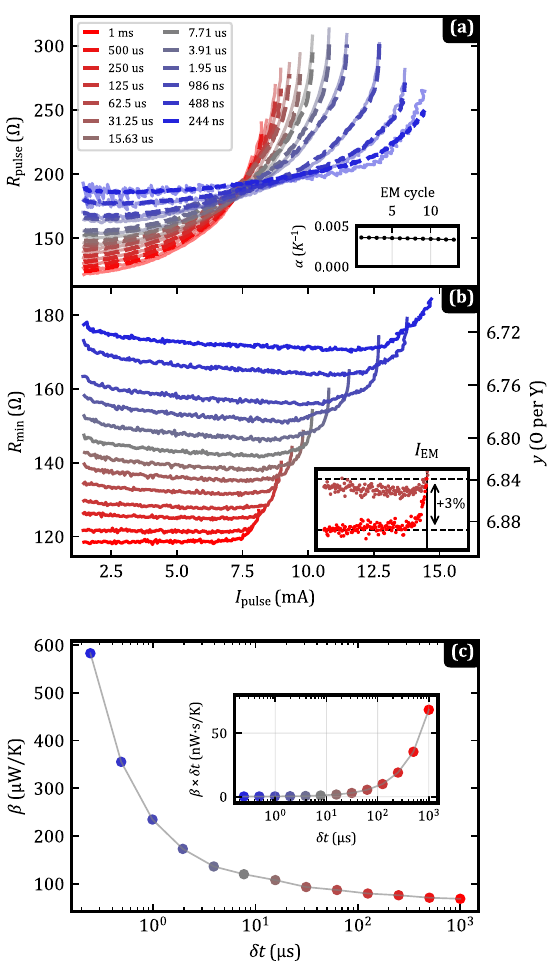}
    \caption{
    \label{fig:EM_curves}
    Consecutive resistance curves for a pulse sweep from long (\SI{1}{ms}) to short (\SI{244}{ns}) pulses in the same sample.
    Panel (a) displays the evolution of $R_{\text{pulse}}$. The inset shows the evolution of the resistivity temperature coefficient $\alpha$ after each electromigration cycle, again derived from the data of \cite{Semba2001}.
    Panel (b) shows the evolution of $R_{\text{min}}$. The gradual increase of $R_{\text{min}}$ with decreasing pulse length arises because all measurements are performed on the same sample. A slight decrease of $R_{\text{min}}$ is observed before entering the electromigration regime. 
    The right axis converts the resistance change into oxygen content using data from \cite{Semba2001}. 
    The inset defines the electromigration current $I_{\text{EM}}$ as the value where $R_{\text{min}}$ has increased by 3\%. The final resistance reached at a given pulse width becomes the starting resistance for the next sweep.
    Panel (c) shows the empirical $\beta$ coefficient as a function of the pulse length. The inset displays the same data as $\beta \delta t$ vs $\delta t$.
}
\end{figure}

Figure~\ref{fig:EM_curves} shows the resistance of the device obtained (a) during the pulse ($R_{\text{pulse}}$) and (b) after the pulse ($R_{\text{min}}$) as a function of the pulsed current amplitude ($I_{\text{pulse}}$) for several pulse lengths. In this case, the first pulse length tested is the longest $\delta t=1$ ms, which is sequentially followed by progressively shorter pulses. The criterion to determine $I_{\text{EM}}$ is based on $R_{\text{min}}$ as illustrated in the inset of panel (b) for the pristine optimally doped YBCO sample. The final $R_{\text{min}}$ value of the electromigration curve for a given pulse length represents the starting resistance value for the next electromigration run using a shorter pulse length, exemplified by the arrow in Fig.~\ref{fig:EM_curves} (b) (curves shift upwards in resistance). Initially, all $R_{\text{min}}(I_{\text{pulse}})$ measurements at low $I_{\text{pulse}}$ values start with a flat slope, which evolves towards a negative slope as the sample is more and more subject to electrical stress. The reason for this trend lies in the fact that the sample is progressively depleted in oxygen content while simultaneously driven to a state of increased oxygen disorder. In other words, the starting state of the sample consists of a somewhat disordered oxygen distribution. At low $I_{\text{pulse}}$, the current is insufficient to electromigrate, but the accompanying heating (thermomigration) stimulates the diffusion of oxygen and the ordering of the oxygen in the Cu-O chains \cite{MARKOWITSCH1996187}, which manifests itself in an improvement of the conductive properties of the sample (i.e. an initial resistance drop) \cite{Marinkovic2023}. In these cases, the 3\% increase of resistance used to define $I_{\text{EM}}$ is taken with respect to the $R_{\text{min}}(I_{\text{pulse}}=0)$.

Unlike $R_{\text{min}}$ curves, the $R_{\text{pulse}}(I_{\text{pulse}})$ response shown in Fig.~\ref{fig:EM_curves}(a) is affected by Joule heating, making it less straightforward to identify the onset of electromigration. It is possible to estimate the shape of the $R_{\text{pulse}}(I_{\text{pulse}})$ curve based on two hypotheses. Firstly, let us assume a linear temperature dependence $ R(T) = R_0 ( 1 + \alpha \Delta T)$, where $\Delta T = T - T_0$ and $R_0 = R(T_0)$. This linear relation only holds for optimally doped YBCO or for not too large $\Delta T$ values. When YBCO undergoes electromigration and the oxygen content locally decreases, the functional dependence becomes moderately sublinear \cite{Semba2001}. Secondly, assuming a steady state, the Joule heating $\mathcal{P} = R_{\text{pulse}} I_{\text{pulse}}^2$ is dissipated in the constriction and evacuated out of the device towards the substrate. Further, considering that the heat transfer coefficients and material parameters do not vary much with temperature in the measured temperature range, the heat flux can be approximated as $\dot Q = \beta \Delta T$, where $\beta$ is a factor taking into account the geometric and thermal parameters. The steady-state heat balance can be expressed as $R_{\text{pulse}} I_{\text{pulse}}^2 = \beta \Delta T$, which leads to,

\begin{equation}
R_{\text{pulse}}(I_{\text{pulse}}) = R_0  \left(1 + \frac{\alpha R_0 I_{\text{pulse}}^2}{\beta - \alpha R_0 I_{\text{pulse}}^2}\right).
\label{eqn:R_pls_fit}
\end{equation}

Note that $R_{\text{pulse}}(I_{\text{pulse}})$ does not follow a simple quadratic dependence. In principle, Eq.~(\ref{eqn:R_pls_fit}) contains two fitting parameters, $\alpha$ and $\beta$. However, based on the dependence of the resistance on the temperature and oxygen content, $R(T,y)$, from Ref.~\cite{Semba2001}, the oxygen content $y=7-\delta$ can be inferred from the resistivity change, which in turn allows us to estimate $\alpha$. This approach permits us to associate the oxygen content corresponding to the $R_{\text{min}}$ values, as shown in the right ordinate axis of Fig.~\ref{fig:EM_curves}(b), and showing that after 13 electromigration runs, the oxygen content has decreased from $y \approx 6.9$ to $y \approx 6.7$. In this range of oxygen contents, the $\alpha$ parameter remain rather constant as shown in the inset of panel (a). Consequently, $\alpha$ is treated as a known parameter, and only $\beta$ remains as a free fitting parameter.

Using Eq.~(\ref{eqn:R_pls_fit}), the data in Fig.~\ref{fig:EM_curves}(a) were fitted as shown by the dotted lines, with $R_0 = R_{\text{min}}$, the above-mentioned strategy to fix $\alpha$, and $\beta$ as the only free parameter. The resulting fit gives $\beta \approx 43$~µW K$^{-1}$ for the 1 ms pulse, increasing progressively to $\beta \approx 475$~µW K$^{-1}$ for the 244 ns pulse as the pulse length decreases.
Figure~\ref{fig:EM_curves}(c) shows the dependence of $\beta$ on $\delta t$. At high $\delta t$, $\beta$ exhibits a low variation. However, for $\delta t < 10$ µs, $\beta$ increases as $\delta t$ decreases. This behavior can be associated with the breakdown of the steady-state hypothesis. Indeed, at short pulse lengths the system does not reach the steady state, then $R_{\text{pulse}}I_{\text{pulse}}^2 \approx C dT/dt \propto \beta \Delta T$ with $C$ the specific heat of the substrate, and therefore $\beta \propto 1/\delta t$. This is indeed confirmed by the rather constant value of $\beta \delta t$ for low $\delta t$, as shown in the inset of Figure~\ref{fig:EM_curves}(c).

\begin{figure}[!ht]
    \centering
    \includegraphics[width=\linewidth]{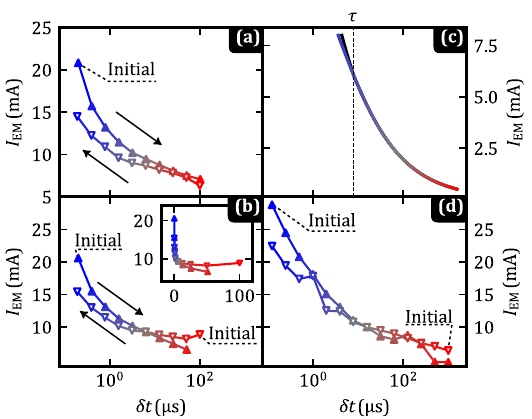}
    \caption{Panel (a) and (b): Dependence of the electromigration current on the pulse length for different scanning orders: long-to-short ($\bigtriangledown$) or short-to-long pulses ($\bigtriangleup$)).
    Panel (a) shows measurements on a single sample during cycling $\delta t$ from short to long pulse lengths and back, whereas panel (b) shows the results for two independent samples.
    (c) Numerical estimation of the electromigration current as a function of the pulse length.
    (d) Dependence of the electromigration current on the pulse length, obtained through finite element simulations.
    }
    \label{fig:I_EM-w_pulse}
\end{figure}

Figure~\ref{fig:I_EM-w_pulse} summarizes the obtained $I_{\text{EM}}$ values as a function of the pulse length $\delta t$. Panel (a) shows the evolution of $I_{\text{EM}}$ when starting from a 200-ns-long pulse, then increasing the pulse length up to 100 ms and subsequently decreasing again down to 200 ns. The irreversible response observed when comparing decreasing and increasing pulse lengths is a direct consequence of the fact that the measurements are made consecutively, and therefore, the sample progressively degrades. As a result, the resistance increases, which then favors electromigration at lower currents due to more pronounced Joule heating. For the sake of comparison, Fig.~\ref{fig:I_EM-w_pulse}(b) shows the response of two different samples subjected to consecutive electromigration processes, one decreasing pulse length and the other increasing pulse length. In this case, the curves approach each other, as expected due to less cumulated irreversibilities. Note that these two curves cross over as a consequence of the fact that at the minimal and maximal $\delta t$ points, respectively, one corresponds to the pristine state, whereas the other has been exposed to cumulated damage. The response $I_{\text{EM}}(\delta t)$ obtained from a set of identical samples would start and finish at the points labeled ``initial" and lie slightly above the two curves. In order to illustrate the dramatic increase of $I_{\text{pulse}}(\delta t)$ as $\delta t$ decreases, in the inset of Fig.~\ref{fig:I_EM-w_pulse}(b), we show the same data as in the main panel, but in a linear $\delta t$ scale.  

The trend observed in the experimental data presented in Fig.~\ref{fig:I_EM-w_pulse}(a,b) can be captured by a simple model. Let us start from the empirical Black's law stating that the mean-time to failure of a device due to electromigration is $MTF=A e^{U/k_BT}J^{-n}$, with $A$ a material and geometrical constant, $U$ is the activation energy for oxygen diffusion, $k_B$ is the Boltzmann constant, $J$ is the current density through the device, and $n \approx 2$ an exponent depending on the electromigration regime \cite{lloyd2002electromigration,hoffmann_vogel_electromigration_2017}. For a given pulse length $\delta t$, the device under consideration fails when the current through it reaches $I_{\text{EM}}$. In other words, for $J=I_{\text{EM}}/S$ with $S$ the cross section of the narrow bridge, and $MTF=\delta t$. From this, it follows 

\begin{equation}
\label{eqn:black}
I_{\text{EM}}^2=A S^2\frac{e^{U/k_BT_{\text{EM}}}}{\delta t} .
\end{equation}

This equation, together with 

\begin{equation}
T_{\text{EM}} = T_0 + \frac{R_0 I_{\text{EM}}^2}{\beta - \alpha R_0 I_{\text{EM}}^2},
\label{eqn:T_EM}
\end{equation}

\noindent results in a system of coupled equations permitting the numerical evaluation of $I_{\text{EM}}(\delta t)$. The result of this calculation is presented in Fig.~\ref{fig:I_EM-w_pulse}(c) for a constant $A$. It is important to point out that the analytical expression in Eq.~(\ref{eqn:T_EM}) contains a denominator $\beta - \alpha R_0 I_{\text{EM}}^2$ which leads to a nonphysical prediction if it approaches zero. Therefore, the model is applicable only for current values below the threshold current $I_{\text{EM}} = \sqrt{\beta/(\alpha R_0)}$.

Let us now estimate the maximum local temperature during pulsed electromigration. There are different approaches to obtain this estimation. The simplest method is to use the sample itself as a thermometer. Indeed, by measuring $R_{\text{pulse}}$, $R_{\text{min}}$, and knowing $\alpha$ we can estimate  $T_{\text{EM}}$ from Eq.~(\ref{eqn:rho_law}),

\begin{equation}
T_{\text{EM}}=T_0+\frac{R_{\text{pulse}}-R_{\text{min}}}{\alpha R_{\text{min}}}.
\label{eqn:sample_thermometer}
\end{equation}

\begin{figure}[!ht]
    \centering
    \includegraphics[width=\linewidth]{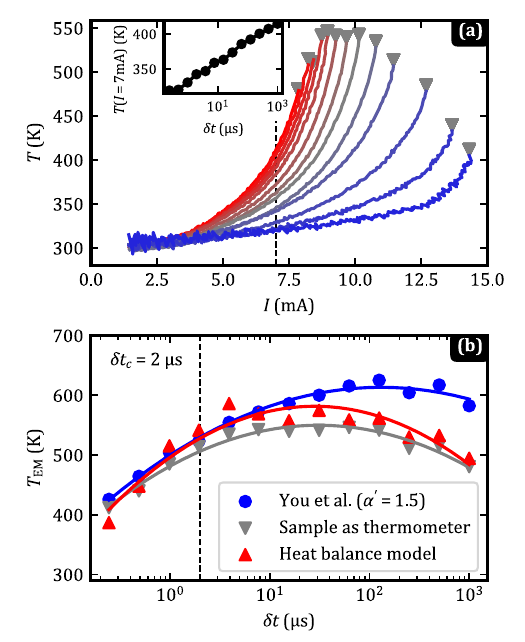}
    \caption{(a) Estimation of the central constriction temperature using Eq.~(\ref{eqn:sample_thermometer}) as a function of the pulsed current. Each curve has been colored according to the pulse length, as in Fig.~\ref{fig:EM_curves}). The inset gives the temperature for a fixed current pulse amplitude (7 mA). It can be seen that longer pulses result in higher temperatures. 
    (b) Comparison of different models to assess the temperature during electromigration for varying pulse lengths. 
    Models: 
    You et al.\cite{you_analytic_2006} (Eq.~(\ref{eqn:you}) with $\alpha = 1$), 
    sample as a thermometer (Eq.~(\ref{eqn:sample_thermometer})), 
    heat balance model (Eq.~(\ref{eqn:T_EM})).
    The vertical line at $\delta t_c = 2$ µs indicates the critical time above which the You et al. model is expected to lose validity. 
    }
    \label{fig:4:T_EM}
\end{figure}

Fig.~\ref{fig:4:T_EM}(a) shows the so-predicted temperature evolution at the central constriction during the electromigration process. The color code related to pulse length is identical to the one in Fig.~\ref{fig:EM_curves}. The maximum temperature at the onset of electromigration is indicated with a gray triangular symbol and shows a non-monotonic dependence on $\delta t$. The inset shows the temperature at the constriction for a fixed pulse current (7 mA), showing that for a fixed current amplitude, longer pulses lead to higher temperatures. Fig.~\ref{fig:4:T_EM}(b) reproduces with gray triangles, the $T_{EM}$ vs $\delta t$ curve obtained from Eq.~(\ref{eqn:sample_thermometer}). Although the derivation of this equation assumes a linear $R(T)$ relationship and uniform heating, it has the advantage of not requiring knowledge of the sample and substrate thermal parameters. An alternative approach is to use the phenomenological model described by Eq.~(\ref{eqn:black}) and Eq.~(\ref{eqn:T_EM}) and numerically extract $T_{\text{EM}}(\delta t)$. This is shown in Fig.~\ref{fig:4:T_EM}(b) with red triangles, assuming $\beta(\delta t)$ to follow the dependence depicted in Fig. 2(c). 

It is interesting to compare these estimations with the analytical expression for the temperature of a current-heated nanowire on a three-dimensional substrate as a function of time, as described in Ref.~\cite{you_analytic_2006,fangohr_joule_2011},

\begin{equation}
T_{\text{EM}} = T_0+\frac{R_{\text{pulse}}I_{\text{EM}}^2}{\pi l k} \operatorname{arcsinh} \left( \frac{2\sqrt{k\delta t/\rho_m C}}{\alpha' w}\right),
\label{eqn:you}
\end{equation}

\noindent where $k=11.7$ WK$^{-1}$m$^{-1}$ is the thermal conductivity of the substrate, $\rho_m=6520$ kgm$^{-3}$ is the density of the substrate, $C=450$ Jkg$^{-1}$K$^{-1}$ is the specific heat of the substrate \cite{Collienne2022}, $w=3$ µm is the width of the wire, $l=5$ µm its length, and $\alpha' = 1.5$ an adjustable parameter. This estimation is presented in Fig.~\ref{fig:4:T_EM}(b) with blue circles.

It is important to note that Eq.~(\ref{eqn:you}) applies to infinitely long nanowires on a semi-infinite substrate. The associated model does not allow for a
temperature variation within the nanowire, and hence, there is translational invariance along the direction of the wire. As detailed in Ref.~\cite{fangohr_joule_2011}, the model leading to Eq.~(\ref{eqn:you}) assumes that the heat front forms
a half-cylinder whose axis is aligned with the wire. As long as the diameter of this half cylinder is small relative to the segment length, the wire appears locally to be infinitely long, and the heat front propagates in a direction perpendicular to the wire. In our case, this assumption is valid only for times shorter than $\delta t_c \approx (0.5l)^2/D_s \approx 2$ µs, where $D_s=k/\rho_mC= 3.8$ mm$^2$s$^{-1}$ is the thermal diffusivity of the substrate. For pulse lengths $\delta t > \delta t_c$, Eq.~(\ref{eqn:you}) tends to overestimate the temperature of the wire, thus explaining the observed slightly different trend with respect to the previous estimations.

The $T(\delta t)$ curves presented in Fig.~\ref{fig:4:T_EM}(b) indicate the presence of two regimes. For $\delta t > 10$ µs, the sample reaches the steady state, and since $I_{EM}$ increases when decreasing $\delta t$, the input power $R_{pulse}I_{EM}^2$ increases and so does the temperature. For $\delta t < 10$ µs, the system remains in the transitory regime, and the sample has no time to heat up. Therefore, shorter pulses result in lower temperatures. In other words, the thermal steady-state assumption is no longer valid below $\delta t = 10$ µs. For this pulse length, the thermal healing length $L_{th}=\sqrt{D_s \delta t} \approx 6$ µm, which nearly corresponds to the length of the constriction. Note that the oxygen diffusion length $L_{\text{diff}, O_2} = \sqrt{D_0\ \text{exp}(-E_a/k_b T)\ \delta t}$ decreases monotonously as pulse length decreases, and does not exceed a few nm, even for the longest pulses.

\section{Conclusion}

We have investigated the onset of the electromigration process in YBCO microbridges under current/voltage pulses spanning pulse lengths $\delta t$ over four orders of magnitude and in the limit of very low duty cycle. We show that pulses longer than 10 µs allow us to reach a thermal steady state, whereas shorter pulses require a full description of the transitory regime. In line with this, the onset current for triggering electromigration exhibits a faster increase as $\delta t$ decreases below 10 µs, since the electromigration process is less assisted by thermal excitations. In order to corroborate this hypothesis, we estimated the maximum temperature reached during the pulsed electromigration as a function of pulse length based on several qualitative, analytical, and numerical models. These findings point to the benefits of using sub-µs current pulses in YBCO circuits to ensure that the electromigration current largely exceeds the critical current of YBCO microwires \cite{Baumans2019}, thereby permitting safe access to the dissipative state without detrimental consequences.    

\section{Acknowledgements}
The authors acknowledge support from Fonds de la Recherche Scientifique - FRS-FNRS, the European COST (European Cooperation in Science and Technology) Action SUPERQUMAP (CA21144), Polytopo (CA23134) and the Spanish Ministry of Science and Innovation MCIN/AEI/10.13039/501100011033/ through the “Severo Ochoa” Programme for Centres of Excellence MaTrans CEX2023-001263-S, HTSUPERFUN PID2021-124680OB-I00 funded by ERDF, “A way of making Europe”, and HTS-4ICT PID2024-156025OB-I00. The Spanish Nanolito networking project (RED2022-134096-T) and Catalan government (2021-SGR-00440) are also acknowledged.

F.\ E. and B.\ A acknowledge support from Aktion Osterreich-Slowakei, AÖSK-Initiativprojektförderung der Aktion (Grant No. 2025-03-15-004)
D.\ S.\ acknowledges support from FRS-FNRS Research Fellowship FRIA.
S.\ M.\ acknowledges support from FRS-FNRS Research Fellowship ASP (Grant No. 1.A.320.21F). 

This research was funded in whole or in part by the Austrian Science Fund (FWF) (Grant DOI: 10.55776/I4865). For the purpose of open access, the authors have applied a CC BY public copyright license to any Author Accepted Manuscript version arising from this submission. Finally, Aktion Österreich-Slowakei, AÖSK-Stipendien für Postdoktoranden (No. MPC-2023-07038) and funding from Slovak Research and Development Agency (No. APVV-21-0231) are gratefully acknowledged.

\section{Annex}
\subsection{Numerical simulation details}
\label{sec:app_mesh}
\begin{figure}[!ht]
    \centering
    \includegraphics[width=\linewidth]{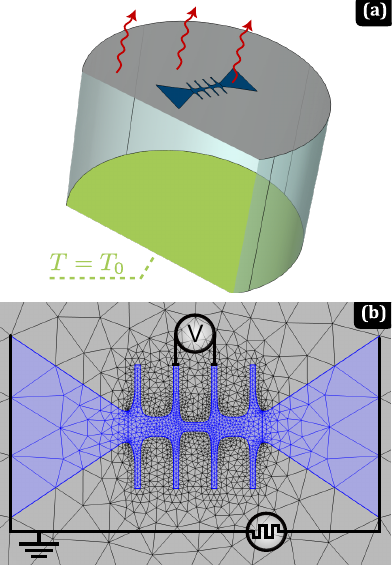}
    \caption{(a) Simulation domain: The back side of the substrate (green) is kept at $T_0$, while the top surfaces (substrate and YBCO film) can also exchange heat via convection at the top). (b) Depiction of the mesh used in the finite element simulations. The blue domain is the YBCO film, in which the (i) heat transfer with Joule heating, (ii) electrokinetics and (iii) electromigration is solved. The leftmost side is kept grounded, whereas the right side is subjected to the current pulses. The difference in electric potential between the two central vertical leads is used to compute the four-point resistance. The gray domain is a part of the substrate, in which only heat transfer is solved for.}
    \label{fig:app_mesh}
\end{figure}

The simulation domain consists of a three-dimensional cylinder as substrate, on top of which the two-dimensional YBCO film is placed, as illustrated in Fig.~\ref{fig:app_mesh}(a).
The following equations are solved in a fully coupled transient study:
(i) The heat equation:
\begin{equation}
    \rho_m C \frac{\partial T}{\partial t} = \nabla . (k \nabla T) + \dot{q}
    - q_{\text{out}},
\end{equation}
where $\dot{q}= J^2 \rho(T, \delta)$ is the heat source resulting from the Joule effect, and $q_{\text{out}}$ is the heat exchanged via natural convection between the top side of the simulation domain (substrate plus YBCO bridge) and the surroundings. The governing equation is given by $q_{\text{out}} = h(T - T_0)$, where $h$ is a convective heat transfer coefficient (initialized to $5\, \mathrm{Wm^{-2}K^{-1}}$ and subsequently calibrated following the method described in Sec.~\ref{sec:3}).
Furthermore, a Dirichlet boundary condition is applied at the back side of the substrate, ensuring $T=T_0$.
The electrical connections are depicted in Fig.~\ref{fig:app_mesh}(b).
(ii) Electrokinetics: To compute the power dissipated by Joule heating as well as the driving forces for electromigration, the Laplace equation is solved, using a temperature and oxygen content dependent resistivity, whose expression has been given in Eq.~(\ref{eqn:rho_law}). 
(iii) Electromigration: Oxygen dynamics is obtained through the mathematical framework described in Ref.~\cite{hoffmann_vogel_electromigration_2017},
\begin{equation}
\frac{\partial C_O}{\partial t} = \mathbf{\nabla} . D(T) \left(\nabla C_O -  \frac{C_O e \rho}{k_B T} z^* \mathbf{J}\right)
\end{equation}
where $C_O$ is the oxygen concentration, $D(T)=D_0\exp(-E_a/(k_\mathrm{B}T))$ is the temperature dependent oxygen diffusion constant and $E_a \approx$ 0.39 eV the associated activation energy. Dirichlet boundary conditions are placed on the current ground and current source terminals to ensure a constant oxygen concentration at these locations.

\subsection{Comparison of the two pulse-generation setups}
\label{sec:app_setup}
\begin{figure}
    \centering
    \includegraphics[width=\linewidth]{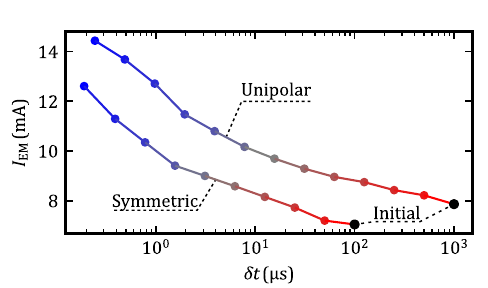}
    \caption{Comparison of measurements performed with the two pulse-generator setups under comparable conditions. Both measurements start from a fresh sample and are performed in the decreasing pulse width direction.}
    \label{fig:app_setup}
\end{figure}

To assess whether the choice of pulse generator influences the measured current-induced oxygen migration, control experiments were performed using both pulse-generation setups under as similar conditions as possible. In particular, the measurements were compared for overlapping ranges of pulse duration and current amplitude. Figure~\ref{fig:app_setup} emphasizes that both setups yielded the same qualitative behavior and comparable trends. This confirms that the reported results are intrinsic to the sample response and not caused by setup-specific artifacts. It has to be noted however, that the electromigration current is slightly higher in the case of the unipolar pulse generator. This can be explained in terms of sample-to-sample variability: $R_0$ was about 20\% higher for the symmetric pulse generator, leading to a lower onset of electromigration due to more power dissipation during the pulse.

\section{References}

\renewcommand{\bibsection}{}

\bibliography{refs.bib}

\end{document}